\documentclass[12pt]{article}

\usepackage{amsmath,amssymb,graphicx,color,array}
\usepackage[pdftex]{hyperref}
\numberwithin{equation}{section}

\evensidemargin \oddsidemargin

\begin{document}

\newcommand{\story}{\vspace{5mm} \noindent $\spadesuit$ }

\begin{titlepage}

\renewcommand{\thefootnote}{\fnsymbol{footnote}}


\begin{flushright}
\end{flushright}

\vspace{15mm}
\baselineskip 9mm
\begin{center}
  {\Large \bf  Probing Smearing Effect by Point-Like Graviton \\
  in Plane-Wave Matrix Model}
\end{center}

\baselineskip 6mm
\vspace{10mm}
\begin{center}
  Bum-Hoon Lee\footnote{\tt bhl@sogang.ac.kr}, and
  Siyoung Nam\footnote{\tt stringphy@gmail.com}
  \\[3mm]
  {\sl Department of Physics 
    and Center for Quantum Spacetime (CQUeST)\\
    Sogang University, C.P.O. Box 1142, Seoul 100-611, South Korea}
  \\[10mm]
  Hyeonjoon Shin\footnote{\tt hshin@sogang.ac.kr} 
  \\[3mm] 
  {\sl Center for Quantum Spacetime (CQUeST)\\ 
    Sogang University, Seoul
    121-742, South Korea }
  \\[3mm]
\end{center}

\thispagestyle{empty}

\vfill
\begin{center}
{\bf Abstract}
\end{center}
\noindent
We investigate the interaction between flat membrane and point-like
graviton in the plane-wave matrix model.  The one-loop effective
potential in the large distance limit is computed and is shown to be
of $r^{-3}$ type where $r$ is the distance between two objects.  This
type of interaction has been interpreted as the one incorporating the
smearing effect due to the configuration of flat membrane in
plane-wave background.  Our result supports this interpretation and
provides one more evidence about it.
\\ [5mm]
Keywords : pp-wave, BMN matrix model, membrane
\\

\vspace{5mm}
\end{titlepage}

\baselineskip 6.6mm
\renewcommand{\thefootnote}{\arabic{footnote}}
\setcounter{footnote}{0}
\setcounter{page}{1}

\section{Introduction}

The plane-wave or BMN matrix model \cite{Berenstein:2002jq} is a model
for the microscopic description of M-theory in the so called
plane-wave background in the framework of the discrete light cone
quantization (DLCQ).  The plane-wave background
\cite{Kowalski-Glikman:1984wv} is $SO(3) \times SO(6)$ symmetric and
given by
\begin{align}
ds^2 &= -2 dx^+ dx^- 
    - \left( \sum^3_{i=1} \left( \frac{\mu}{3} \right)^2 (x^i)^2
            +\sum^9_{a=4} \left( \frac{\mu}{6} \right)^2 (x^a)^2
      \right) (dx^+)^2
    + \sum^9_{I=1} (dx^I)^2 ~,
     \notag \\
F_{+123} &= \mu ~,
\label{pp}
\end{align}
with the index notation $I=(i,a)$. This background is maximally
supersymmetric and obtained by taking the Penrose limit to the
eleven-dimensional AdS type geometries \cite{Blau:2002dy}.

From the structural point of view, the plane-wave matrix model is a
mass deformation of the matrix model in flat spacetime, the BFSS
matrix model \cite{Banks:1997vh}.  Compared to the BFSS matrix model,
one distinguished feature of the plane-wave matrix model is that the
supersymmetric fuzzy sphere membrane with finite size appears from the
vacuum structure \cite{Berenstein:2002jq,Dasgupta:2002hx}.  Although
it is a configuration of membrane, it has been interpreted as a
graviton, or more precisely a giant graviton because it has a size.
The presence of the fuzzy sphere membrane has led to a lot of works
studying its nature from various viewpoints
\cite{Dasgupta:2002hx}-\cite{Bak:2005jh}.  In the study of dynamical
aspect, it has been shown that the fuzzy sphere behaves indeed like a
graviton, and evidences about its interpretation as a giant graviton
have been accumulated \cite{Lee:2003kf}-\cite{Shin:2008ep}.  The
thermodynamical aspect of fuzzy sphere has also been considered in
\cite{Shin:2004ms}-\cite{Kawahara:2007nw}.  Upon a proper circle
compactification, the plane-wave matrix model leads to the matrix
string theory, which is related in the infrared limit to the free
string theory in ten-dimensional plane wave background
\cite{Sugiyama:2002tf}-\cite{Lozano:2006jr}.  This string theory
contains fuzzy spheres in its spectrum, whose various aspects also
have been studied \cite{Hyun:2003ks}-\cite{Das:2005vd}.

Among the studies on fuzzy sphere, an interesting result has been
obtained in the investigation of the interaction between fuzzy sphere
and flat membrane by one of the present authors \cite{Shin:2008ep}.
At one-loop level, the effective potential has been calculated in the
large $r$ limit, where $r$ is the distance between two objects.
Because the fuzzy sphere can be regarded as a point-like object at
large $r$, the leading interaction potential was expected to be of
$r^{-5}$ type based on the result from the BFSS matrix model
\cite{Aharony:1996bh}.  However, the leading interaction has turned
out to be of $r^{-3}$ type rather than $r^{-5}$ type.  The
interpretation for this unexpected result was as follows: If the
supersymmetric flat membrane is placed in the $SO(6)$ symmetric space
of the plane-wave then it spans and spins in four dimensional subspace
basically due to the nature of plane-wave background.  As a result,
two more extra dimensions are required for its configuration.  This
fact is reflected in the interaction potential as the delocalization
or smearing effect.  Actually, the smearing effect has been already
reported in the supergravity side \cite{Bain:2002nq,Mas:2003uk}.  In
the plane-wave background, it has been observed that some supergravity
solutions show the delocalization or smearing of branes in some
directions.

Although the interpretation of the interaction in terms of the
smearing effect is interesting, it has been given from just one
example and one may wonder whether we can give the same interpretation
in other cases involving the flat membrane.  In this paper, we will
study the interaction from another configuration for the purpose of
checking the previous interpretation, and give one more evidence about
the smearing effect due to the flat membrane in the plane-wave matrix
model.  The configuration we will take is composed of one point-like
graviton and one flat membrane, each of which is supersymmetric.  Two
objects are kept apart on a plane in the $SO(3)$ symmetric space.  We
note that the situation is different from the previous case
\cite{Shin:2008ep} where one fuzzy sphere is separated from one flat
membrane on a plane in the $SO(6)$ symmetric space.  Actually, this
gives the reason why we take the point-like graviton as the object
interacting with the flat membrane.  Basically, the interaction
between supersymmetric objects is our concern.  Contrary to the fuzzy
sphere, the point-like graviton can be supersymmetric even if it has a
motion in the $SO(3)$ symmetric space \cite{Park:2002cba}.

The organization of this paper is as follows.  In the next section, we
will give a brief introduction to the plane-wave matrix model.  In
Sec.~\ref{bgc}, the background configuration composed of flat membrane
and point-like graviton is presented.  In Sec.~\ref{1-loop-path-int},
the formal one-loop path integration of the plane-wave matrix model
around the background configuration of Sec.~\ref{bgc} is performed.
From the result of path integration, the one-loop effective potential
is obtained in Sec.~\ref{epotential}.  Finally, the conclusion follows
in Sec.~\ref{finalsec}.

\section{Plane-wave matrix model}

The plane-wave matrix model is basically composed of two parts.  One
part is the usual matrix model based on eleven-dimensional flat
space-time, that is, the flat space matrix model \cite{Banks:1997vh},
and another is a set of terms depending on $\mu$ and reflecting the
structure of the maximally supersymmetric eleven dimensional
plane-wave background, Eq. (\ref{pp}).  Its action is
\begin{align}
S_{pp} = &  \int dt \mathrm{Tr} 
\left( \frac{1}{2} D_t X^I D_t X^I + \frac{1}{4} ( [ X^I, X^J] )^2
      + i \Theta^\dagger D_t \Theta 
      -  \Theta^\dagger \gamma^I [ \Theta, X^I ]
\right.
  \notag \\
  &  \left. 
      -\frac{1}{2} \left( \frac{\mu}{3} \right)^2 (X^i)^2
      -\frac{1}{2} \left( \frac{\mu}{6} \right)^2 (X^a)^2
      - i \frac{\mu}{3} \epsilon^{ijk} X^i X^j X^k
      - i \frac{\mu}{4} \Theta^\dagger \gamma^{123} \Theta
\right) \, ,
\label{pp-action}
\end{align}
where $D_t$ is the covariant derivative with the gauge field $A$,
\begin{equation}
D_t = \partial_t - i [A, \: ] ~,
\end{equation}
and $\gamma^I$ is the $16 \times 16$ $SO(9)$ gamma matrices.

In matrix model, various objects, like branes and graviton, are
realized by the classical solutions of the equations of motion for the
matrix field, which are derived from Eq.~(\ref{pp-action}) as follows:
\begin{align}
\ddot{X}^i &=
   - [[ X^i, X^I],X^I] - \left( \frac{\mu}{3} \right)^2 X^i
   - i \mu \epsilon^{ijk} X^j X^k ~, 
                 \nonumber \\
\ddot{X}^a &=
    - [[ X^a, X^I],X^I] - \left( \frac{\mu}{6} \right)^2 X^a ~,
\label{eom}
\end{align}
where the over dot implies the time derivative $\partial_t$.  Here,
since the object that we are concerned about is purely bosonic, only
the equations of motion for the bosonic field have been presented.
For given objects, the dynamics between them is studied by expanding
the matrix model action around the corresponding classical solution
and performing the path integration.  Let us denote the classical
solution or the background configuration by $B^I$, and split the
matrix quantities into as follows:
\begin{equation}
X^I = B^I + Y^I \, , \quad A=0+A \, , \quad \Theta = 0 + \Psi \, .
\label{cl+qu}
\end{equation}
Then $Y^I$, $A$ and $\Psi$ are the quantum fluctuations around the
background configuration, which are the fields subject to the path
integration.  We note that the gauge field may also have non-trivial
classical configuration.  However, it is simply set to zero in this
paper because the objects we are interested in do not generate any
background gauge field.

In taking into account the quantum fluctuations, we should recall that
the matrix model itself is a gauge theory.  This implies that the
gauge fixing condition should be specified before proceed further.  In
this paper, we take the background field gauge which is usually chosen
in the matrix model calculation,
\begin{equation}
D_\mu^{\rm bg} A^\mu_{\rm qu} \equiv
D_t A + i [ B^I, X^I ] = 0 ~.
\label{bg-gauge}
\end{equation}
Then the corresponding gauge-fixing $S_\mathrm{GF}$ and Faddeev-Popov
ghost $S_\mathrm{FP}$ terms are given by
\begin{equation}
S_\mathrm{GF} + S_\mathrm{FP} =  \int\!dt \,{\rm Tr}
  \left(
      -  \frac{1}{2} (D_\mu^{\rm bg} A^\mu_{\rm qu} )^2 
      -  \bar{C} \partial_{t} D_t C + [B^I, \bar{C}] [X^I,\,C]
\right) ~.
\label{gf-fp}
\end{equation}

Now by inserting the decomposition of the matrix fields (\ref{cl+qu})
into Eqs.~(\ref{pp-action}) and (\ref{gf-fp}), we get the gauge fixed
plane-wave action $S$ $(\equiv S_{pp} + S_\mathrm{GF} +
S_\mathrm{FP})$ expanded around the classical background $B^I$.  The
resulting action is read as
\begin{equation}
S =  S_0 + S_2 + S_3 + S_4 ~,
\end{equation}
where $S_k$ represents the action of order $k$ with respect to the
quantum fluctuations and, for each $k$, its expression is
\begin{align}
S_0 = \int dt \, \mathrm{Tr} \bigg[ \,
&      \frac{1}{2}(\dot{B}^I)^2  
        - \frac{1}{2} \left(\frac{\mu}{3}\right)^2 (B^i)^2 
        - \frac{1}{2} \left(\frac{\mu}{6}\right)^2 (B^a)^2 
        + \frac{1}{4}([B^I,\,B^J])^2
        - i \frac{\mu}{3} \epsilon^{ijk} B^i B^j B^k 
    \bigg] ~,
\notag \\
S_2 = \int dt \, \mathrm{Tr} \bigg[ \,
&       \frac{1}{2} ( \dot{Y}^I)^2 - 2i \dot{B}^I [A, \, Y^I] 
        + \frac{1}{2}([B^I , \, Y^J])^2 
        + [B^I , \, B^J] [Y^I , \, Y^J]
        - i \mu \epsilon^{ijk} B^i Y^j Y^k
\notag \\
&       - \frac{1}{2} \left( \frac{\mu}{3} \right)^2 (Y^i)^2 
        - \frac{1}{2} \left( \frac{\mu}{6} \right)^2 (Y^a)^2 
        + i \Psi^\dagger \dot{\Psi} 
        -  \Psi^\dagger \gamma^I [ \Psi , \, B^I ] 
        -i \frac{\mu}{4} \Psi^\dagger \gamma^{123} \Psi  
\notag \\ 
&       - \frac{1}{2} \dot{A}^2  - \frac{1}{2} ( [B^I , \, A])^2 
        + \dot{\bar{C}} \dot{C} 
        + [B^I , \, \bar{C} ] [ B^I ,\, C] \,
     \bigg] ~,
\notag \\
S_3 = \int dt \, \mathrm{Tr} \bigg[
&       - i\dot{Y}^I [ A , \, Y^I ] - [A , \, B^I] [ A, \, Y^I] 
        + [ B^I , \, Y^J] [Y^I , \, Y^J] 
        +  \Psi^\dagger [A , \, \Psi] 
\notag \\
&       -  \Psi^\dagger \gamma^I [ \Psi , \, Y^I ] 
        - i \frac{\mu}{3} \epsilon^{ijk} Y^i Y^j Y^k
        - i \dot{\bar{C}} [A , \, C] 
        +  [B^I,\, \bar{C} ] [Y^I,\,C]  \,
     \bigg] ~,
\notag \\
S_4 = \int dt \, \mathrm{Tr} \bigg[
&       - \frac{1}{2} ([A,\,Y^I])^2 + \frac{1}{4} ([Y^I,\,Y^J])^2 
     \bigg] ~.
\label{bgaction} 
\end{align}

\section{Background configuration}
\label{bgc}

In this section, we set up the background configuration corresponding
to one flat membrane and one graviton, and discuss about the
perturbation theory around it.

Since the background is composed of two objects, the matrices
representing the background have the $2 \times 2$ block diagonal form
as
\begin{align}
B^I = \begin{pmatrix} B^I_{(1)} & 0 \\ 0 & B^I_{(2)} \end{pmatrix}
\label{bgconfig}
\end{align}
where $B_{(s)}^I$ with $s=1,2$ are $N_s \times N_s$ matrices.  If
$B^I$ are taken to be $N \times N$ matrices, then $N = N_1 + N_2$.

The first object represented by $B^I_{(1)}$ is taken to be the flat
membrane found in \cite{Hyun:2002cm}.  It is 1/8-BPS object, and spans
and spins in four dimensional subspace of the $SO(6)$ symmetric space
as
\begin{gather}
B^4_{(1)} = Q \cos(\mu t/6) \,, \quad B^6_{(1)} = Q \sin(\mu t/6)\,,
\notag \\
B^5_{(1)} = P \cos(\mu t/6)\,, \quad B^7_{(1)} = P \sin(\mu t/6)\,,
\label{bg1}
\end{gather}
where $N_1 \times N_1$ matrices, $Q$ and $P$, satisfy
\begin{align}
[ Q, P ] = i \sigma \, ,
\label{flatcom}
\end{align}
with a small constant parameter $\sigma$.  We note that, in order to
describe the flat membrane properly, the size of the matrix should be
infinite.  In what follows, $N_1$ is thus implicitly taken to be
infinite.  Now, from this somewhat complicated configuration, we see
that, at $t=0$, the flat membrane is placed in $x^4$-$x^5$ plane, and,
as time goes by, one axis along $x^4$ rotates in $x^4$-$x^6$ plane
while another axis along $x^5$ rotates in $x^5$-$x^7$ plane.  At this
point, one may notice that the configuration looks quite strange since
the membrane of infinite size is spinning.  However, this kind of
situation may be understood as the one due to the choice of the
coordinate system for representing the plane-wave background
(\ref{pp}), and may disappear by going to a frame where the membrane
looks like static one.  Indeed, as we will see, it is possible to take
a certain frame where the whole background configuration looks like
static one.

The graviton is the second object $B^I_{(2)}$, which is represented by
$1 \times 1$ matrix and taken to rotate in $x^1$-$x^2$ as follows:
\begin{align}
B^1_{(2)} = r \cos(\mu t/3)  \,, \quad
B^2_{(2)} = r \sin(\mu t/3) \,,
\label{bg2}
\end{align}
which is 1/2-BPS object as shown in \cite{Park:2002cba}.  Thus the
graviton is placed in the transverse space of the flat membrane with
distance $r$.

For the background configuration (\ref{bgconfig}) given above, the
classical value of the action $S_0$ is evaluated as $S_0/T = -
\frac{1}{2} N_1 \sigma^2$ where $T=12 \pi/\mu$ is the period of motion
for the membrane. Here, since the motion is periodic, we have
considered the action per one period.  As for the graviton, the
classical action simply vanishes.

From now on, what we are going to do is the computation of the
one-loop correction to the classical action $S_0$, that is, to the
background, (\ref{bg1}) and (\ref{bg2}), due to the quantum
fluctuations via the path integration of the quadratic action $S_2$,
and obtain the one-loop effective action $\Gamma_\mathrm{eff}$ or the
effective potential $V_\mathrm{eff}$.  But, in order to justify the
one-loop computation, it should be made clear that $S_3$ and $S_4$ of
Eq.~(\ref{bgaction}) can be regarded as perturbations.  For this
purpose, following \cite{Dasgupta:2002hx}, we rescale the fluctuations
and parameters as
\begin{gather}
A   \rightarrow \mu^{-1/2} A   \, , \quad
Y^I \rightarrow \mu^{-1/2} Y^I \, , \quad
C   \rightarrow \mu^{-1/2} C   \, , \quad
\bar{C} \rightarrow \mu^{-1/2} \bar{C} \, ,
\notag \\
r \rightarrow \mu r \, , \quad
t \rightarrow \mu^{-1} t \, , \quad
Q \rightarrow \mu Q \, , \quad
P \rightarrow \mu P \, , \quad
\sigma \rightarrow \mu^2 \sigma \, .
\label{rescale}
\end{gather}
Under this rescaling, the powers of $\mu$ are factored out from the
action $S$ in the background (\ref{bg1}) and (\ref{bg2}) as
\begin{align}
S =  \mu^3 S_0 + S_2 + \mu^{-3/2} S_3 + \mu^{-3} S_4 ~,
\label{ssss}
\end{align}
where $S_0$, $S_2$, $S_3$ and $S_4$ do not have $\mu$ dependence.
Obviously, this form of the action ensures us that, in the large $\mu$
limit, $S_3$ and $S_4$ can be treated as perturbations and the
one-loop computation gives the sensible result.

Based on the structure of (\ref{bgconfig}), we now write the quantum
fluctuations in the $2 \times 2$ block matrix form as follows:
\begin{gather}
A   = \begin{pmatrix}
          0               &   \Phi^0      \\
         \Phi^{0 \dagger} &   0
      \end{pmatrix} ~,~~~
Y^I = \begin{pmatrix}
          0               &   \Phi^I      \\
         \Phi^{I \dagger} &   0
      \end{pmatrix} ~,~~~
\Psi = \begin{pmatrix}
          0             &  \chi \\
         \chi^{\dagger} &  0
       \end{pmatrix} ~,
 \notag \\
C =    \begin{pmatrix}
         0           &  C  \\ 
         C^{\dagger} &  0
       \end{pmatrix} ~,~~~
\bar{C} = \begin{pmatrix}
              0               &  \bar{C}  \\
              \bar{C}^\dagger &  0
           \end{pmatrix} ~.
\label{q-fluct}
\end{gather}
Although we denote the block off-diagonal matrices for the ghosts by
the same symbols with those of the original ghost matrices, there will
be no confusion since $N \times N$ matrices will never appear from now
on.  The reason why the block-diagonal parts are not considered is
that they do not give any effect on the interaction between two
objects at least at the one-loop level.

\section{One-loop quantum fluctuations}
\label{1-loop-path-int}

In this section, we perform the path integration for the quadratic
action, $S_2$, around the classical background (\ref{bgconfig}) with
(\ref{bg1}) and (\ref{bg2}).  We will state only the formal results
whose actual evaluation will be described in the next section.

The quadratic action is largely composed of three decoupled sectors,
which are bosonic, ghost, and fermionic sectors.  In the path
integration of each sector, the integration variables are matrices.
For the actual evaluation of the path integration, it is usually
useful to expand the matrix variables in a suitable matrix basis.
Taking a matrix basis depends on the classical background under
consideration.

For the present case where the flat membrane is involved, the
commutation relation (\ref{flatcom}), the characteristic of the flat
membrane provides a clue for the desired matrix basis.  If we define
\begin{align}
a = \frac{1}{\sqrt{2 \sigma}} (Q+iP) ~, \quad
a^\dagger = \frac{1}{\sqrt{2 \sigma}} (Q-iP)~,
\label{osc-ca}
\end{align}
then they satisfy the commutation relation
\begin{align}
[ a, a^\dagger ] =1 \, ,
\label{comrel}
\end{align}
and can be regarded as the annihilation and creation operators of
simple harmonic oscillator. This fact allows us to express the
fluctuations around the flat membrane in terms of the oscillator
states, on which $a$ and $a^\dagger$ act as
\begin{align}
a | n \rangle = \sqrt{n} | n-1 \rangle \, , \quad
a^\dagger | n \rangle = \sqrt{n+1} | n+1 \rangle \, .
\label{osc-alg}
\end{align}
Because the size of the membrane is given by $N_1$, the oscillator
number $n$ runs from $0$ to $N_1-1$, and hence has the upper bound.
However, we note that actually there is no upper bound for $n$ because
$N_1$ should be infinite for the proper description of the flat
membrane.  So, we take $N_1$ to be infinite from now on.

The off-diagonal blocks of Eq.~(\ref{q-fluct}) are simply vectors
because the point-like graviton background is described by $1\times 1$
matrix.  This fact and above consideration leads us now to take $| n
\rangle$ as the matrix basis for the fluctuations.  Then, in this
matrix basis, each fluctuation matrix has the following type of matrix
mode expansion
\begin{align}
\Phi = \sum_{n=0}^{\infty} \phi_n | n \rangle \,.
\label{decom}
\end{align}
This expansion allows us to reduce the path integration of the matrix
variable to that of the mode $\phi_n$.

\subsection{Bosonic contributions}

We first consider the path integration of bosonic fluctuations
including also the ghost part.

The Lagrangian $L_B$ for the purely bosonic fluctuations is given by
\begin{align}
L_B = \mathrm{Tr}
 \bigg\{ 
&\,
- | \dot{\Phi}^0 |^2 
  + \Phi^{0\dagger} (r^2 +Q^2 + P^2) \Phi^0
\notag \\
&+ | \dot{\Phi}^I |^2 
  - \Phi^{I\dagger} ( r^2 + Q^2+P^2) \Phi^I
  - \frac{1}{3^2} |\Phi^i|^2 
  - \frac{1}{6^2} |\Phi^a|^2 
\notag \\
& - ir \sin(t/3) (\Phi^{3\dagger}\Phi^1
                 -\Phi^{1\dagger}\Phi^3)
  + ir \cos(t/3) (\Phi^{3\dagger}\Phi^2
                 -\Phi^{2\dagger}\Phi^3)				 
\notag \\ 
& + i \frac{2}{3} r \sin (t/3) 
          (\Phi^{0\dagger} \Phi^1 - \Phi^{1\dagger}\Phi^0)
  - i \frac{2}{3} r \cos (t/3) 
          (\Phi^{0\dagger} \Phi^2 - \Phi^{2\dagger}\Phi^0)
\notag \\
& - \frac{i}{3} \sin (t/6)
  \left[
   \Phi^{0 \dagger} ( Q \Phi^4 + P \Phi^5)   
    - ( \Phi^{4\dagger} Q + \Phi^{5 \dagger} P ) \Phi^0 
  \right]
\notag \\
&  + \frac{i}{3} \cos (t/6)
  \left[
   \Phi^{0 \dagger} ( Q \Phi^6 + P \Phi^7 )   
    - (\Phi^{6\dagger} Q + \Phi^{7 \dagger} P) \Phi^0 
  \right]  
\notag \\
&  -2 i \sigma  
  \Big[
    \left(\cos (t/6)\Phi^{4 \dagger} 
      + \sin (t/6) \Phi^{6 \dagger}
    \right)
    \left(\cos (t/6) \Phi^5 
      + \sin (t/6) \Phi^7
    \right)
\notag \\
&   - \left( \cos (t/6) \Phi^{5 \dagger} 
       + \sin (t/6) \Phi^{7 \dagger}
    \right) 
    \left( \cos (t/6)\Phi^4 
         + \sin (t/6) \Phi^6
    \right)
  \Big]   
\bigg\} ~.
\label{olb}
\end{align}
Because of the rotating background, the Lagrangian depends on time
explicitly through trigonometric functions, which makes the path
integration cumbersome.  This kind of explicit time dependence can be
removed by going to a certain frame where the background looks like
static.  In fact, changing frame is natural, since the configuration
in the original frame contains the rotating membrane of infinite size
which is quite strange in physical sense.  Then, in order to move to
the desired frame, we take
\begin{align}
\cos(t/3) \Phi^1 + \sin(t/3) \Phi^2 \rightarrow \Phi^1\,,\quad
-\sin(t/3) \Phi^1 + \cos(t/3) \Phi^2 \rightarrow \Phi^2\, ,
\notag \\
\cos(t/6) \Phi^4 + \sin(t/6) \Phi^6 \rightarrow \Phi^4\,,\quad
-\sin(t/6) \Phi^4 + \cos(t/6) \Phi^6 \rightarrow \Phi^6\,,
\notag \\
\cos(t/6) \Phi^5 + \sin(t/6) \Phi^7 \rightarrow \Phi^5\,,\quad
-\sin(t/6) \Phi^5 + \cos(t/6) \Phi^7 \rightarrow \Phi^7\,.
\end{align}
Under these transformations, the above Lagrangian (\ref{olb}) becomes
\begin{align}
L_B = \mathrm{Tr} \Big[ \,
& - | \dot{\Phi}^0 |^2 
  + \Phi^{0\dagger} (r^2 +Q^2 + P^2) \Phi^0
\notag \\
& + | \dot{\Phi}^I |^2 
  - \Phi^{I\dagger }(r^2 +Q^2+P^2)\Phi^I
  - \frac{1}{3^2} | \Phi^3 |^2 
  - \frac{1}{6^2} | \Phi^8 |^2 - \frac{1}{6^2} | \Phi^9 |^2
\notag \\  
& +\frac{2}{3}
     (\Phi^{1 \dagger} \dot{\Phi}^{2}
      - \Phi^{2 \dagger} \dot{\Phi}^1 ) 
  +\frac{1}{3}
     (\Phi^{4 \dagger} \dot{\Phi}^6
      - \Phi^{6 \dagger} \dot{\Phi}^4 ) 
  +\frac{1}{3}
     (\Phi^{5 \dagger} \dot{\Phi}^7
      - \Phi^{7 \dagger} \dot{\Phi}^5 )
\notag \\
& - ir ( \Phi^{2\dagger} \Phi^3 - \Phi^{3\dagger} \Phi^2 )     
- \frac{i}{3} 
   \Phi^{0 \dagger} (2r \Phi^2
     - Q \Phi^6 - P \Phi^7 )   
\notag \\
& + \frac{i}{3} 
   (2r \Phi^{2\dagger}  
   - \Phi^{6 \dagger}Q - \Phi^{7 \dagger} P) 
   \Phi^0 
  -2 i \sigma  (\Phi^{4 \dagger} \Phi^5 - \Phi^{5 \dagger} \Phi^4 )   
 \, \Big] \, ,
\label{lb}
\end{align}
which is obviously free of trigonometric functions having explicit
time dependence.

The first observation for the Lagrangian (\ref{lb}) is that the matrix
fields $\Phi^8$ and $\Phi^9$ are free and decoupled from other fields.
This means that the path integration for these fields can be carried
out immediately.  If we use the matrix expansion Eq.~(\ref{decom}) for
each of $\Phi^8$ and $\Phi^9$, and the relation
\begin{align}
(Q^2+P^2) | n \rangle = \sigma (2 a^\dagger a +1 ) |n \rangle
=\sigma (2 n +1 ) | n \rangle \,,
\end{align}
which is derived from Eqs.~(\ref{osc-ca}), (\ref{comrel}), and
(\ref{osc-alg}), then the result of the path integration is simply
obtained as
\begin{align}
\prod^\infty_{n=0} {\det}^{-2}
\left( \Delta_n - \frac{1}{6^2} \right) \,,
\label{bdet1}
\end{align}
where we have defined
\begin{align}
\Delta_{n} \equiv - \partial_t^2 - r^2 - \sigma (2n+1) \,.
\label{deln}
\end{align}

Other matrix fields except for $\Phi^8$ and $\Phi^9$ are coupled to
each other, and they should be taken into account as a whole.  Let us
denote the Lagrangian describing them as $\widehat{L}_B$, which is
given by Eq.~(\ref{lb}) with vanishing $\Phi^8$ and $\Phi^9$.  That
is,
\begin{align}
\widehat{L}_B = \left. L_B \right|_{\Phi^8,\Phi^9=0} \,.
\end{align}
When each matrix variable is expanded in terms of Eq.~(\ref{decom}),
we can express this Lagrangian in terms of modes.  By the way, as
already noted in a previous work \cite{Shin:2008ep} done by one of the
present authors, the terms linear in $Q$ and $P$ lead to coupling of
modes with different oscillator number $n$ because $Q$ and $P$ are
linear combinations of the creation and annihilation operators as seen
in Eq.~(\ref{osc-ca}).  In order to avoid such mixing, we follow the
prescription given in \cite{Aharony:1996bh}, and take the unitary
transformation as follows:
\begin{align}
\Phi^\pm \equiv 
   \frac{1}{\sqrt{2}} ( \Phi^4 \pm i \Phi^5 ) \, ,
\quad
\tilde{\Phi}^\pm \equiv 
   \frac{1}{\sqrt{2}} ( \Phi^6 \pm i \Phi^7 ) \,.
\end{align}
Then the terms linear in $Q$ and $P$ become
\begin{align}
& \mathrm{Tr} \left[ 
 \Phi^{0 \dagger} (Q \Phi^6 + P \Phi^7 )   
 -  (\Phi^{6 \dagger}Q + \Phi^{7 \dagger}P ) \Phi^0
\right]
\notag \\
=& \sqrt{\sigma} \mathrm{Tr} \left[
  \Phi^{0 \dagger} ( a \tilde{\Phi}^- +  a^\dagger \tilde{\Phi}^+ )   
 - ( \tilde{\Phi}^{+ \dagger} a + \tilde{\Phi}^{- \dagger} a^\dagger ) 
  \Phi^0 \right] \, .
\end{align}
From this structure, we see that the matrix mode expansions for
$\Phi^\pm$ and $\tilde{\Phi}^\pm$ should be taken as
\begin{align}
\Phi^\pm = \sum_{n=\pm 1}^{\infty} \phi^\pm_n 
| n \mp 1 \rangle \, , \quad
\tilde{\Phi}^\pm = \sum_{n=\pm 1}^{\infty} 
\tilde{\phi}^\pm_n | n \mp 1 \rangle  \, ,
\label{mode}
\end{align}
if $\Phi^0$, $\Phi^1$, $\Phi^2$, and $\Phi^3$ are taken to follow the
expansion of Eq.~(\ref{decom}).  Here, the reason why $\Phi^\pm$ and
$\tilde{\Phi}^\pm$ should have the same type of mode expansion is that
they couple to each other with one time derivative.

In terms of the matrix mode expansions described above, the Lagrangian
is now written without any mixing between different oscillator number
$n$ as
\begin{align}
\widehat{L}_B = \sum_{n=0}^\infty V_n^\dagger M_n V_n \,,
\label{Blag}
\end{align}
where
$ V_n = (
\phi^0_n , \, \phi^1_n , \, \phi^2_n , \, \phi^3_n , \, 
\phi^+_n , \, \tilde{\phi}^+_n , \,
\phi^-_n , \, \tilde{\phi}^-_n )^T $ 
and
\begin{align}
M_n = 
\begin{pmatrix}
-\Delta_n & 0 & -\frac{2i}{3}r & 0 & 0 & 
  \frac{i}{3} \sqrt{\sigma n} & 0 & \frac{i}{3} \sqrt{\sigma (n+1)} \\
0 & \Delta_n & \frac{2}{3} \partial_t & 0 & 0 & 0 & 0 & 0 \\
\frac{2i}{3} r & -\frac{2}{3} \partial_t & \Delta_n & - i r
  & 0 & 0 & 0 & 0 \\
0 & 0 & i r & \Delta_n-\frac{1}{3^2} & 0 & 0 & 0 & 0 \\
0& 0 & 0 & 0 & \Delta_n & \frac{1}{3} \partial_t & 0 & 0  \\
-\frac{i}{3} \sqrt{\sigma n} & 0 & 0 & 0 & -\frac{1}{3} \partial_t 
       & \Delta_n + 2\sigma  & 0 & 0 \\
0 & 0 & 0 & 0 & 0 & 0 & \Delta_n & \frac{1}{3} \partial_t  \\
-\frac{i}{3} \sqrt{\sigma (n+1)} & 0 & 0 & 0 & 0 & 0 & 
-\frac{1}{3} \partial_t &  \Delta_n - 2 \sigma
\end{pmatrix} \, ,
\end{align}
with the $\Delta_n$ defined in Eq.~(\ref{deln}).  One may notice that
the summation over $n$ starts from 0 while the minimum value of $n$ in
the mode expansions of Eq.~(\ref{mode}) is $+1$ or $-1$.  We
illuminate on this. The oscillator number $n$ of $\phi^+_n$ and
$\tilde{\phi}^+_n$ starts from $+1$ while that of $\phi^-_n$ and
$\tilde{\phi}^-_n$ starts from $-1$.  It is easy to see that the modes
$\phi^-_{-1}$ and $\tilde{\phi}^-_{-1}$ are decoupled from other modes
and form a subsystem, because all other modes do not have such
oscillator number.  As for the modes $\phi^+_n$ and
$\tilde{\phi}^+_n$, the absence of them at $n=0$ seems to require an
independent treatment of $M_0$.  However, let us suppose that these
modes were present at the beginning.  Then, the structure of $M_0$
shows that they would be decoupled from other modes and form a
subsystem.  Furthermore, the subsystem is exactly the same with that
composed of $\phi^-_{-1}$ and $\tilde{\phi}^-_{-1}$.  This indicates
that the modes $\phi^-_{-1}$ and $\tilde{\phi}^-_{-1}$ can be
symbolically identified with $\phi^+_0$ and $\tilde{\phi}^+_0$.  More
precisely, $\phi^-_{-1} \rightarrow \tilde{\phi}^+_0$ and
$\tilde{\phi}^-_{-1} \rightarrow \phi^+_0$, which can be inferred from
$M_0$.  After all, all the modes can be taken to have the oscillator
number starting from $n=0$.  We would like to note that similar
situation also appears in the investigation of the interaction between
fuzzy sphere and flat membrane \cite{Shin:2008ep}.

The mode expanded Lagrangian (\ref{Blag}) allows us to evaluate the
path integral as
\begin{align}
\prod^\infty_{n=0} \mathrm{Det}^{-1} M_n \, ,
\end{align}
where $\mathrm{Det}$ involves the matrix determinant as well as the
usual functional one.  Usually, for the one-loop effective action or
potential, the diagonalization of the matrix $M_n$ is preferable.
However, it is not an easy task, basically due to the two constant
terms $\pm 2 \sigma$ appearing in the diagonal elements of the
matrix. Fortunately, we do not need to have the fully factorized form
of the determinant of $M_n$, since we are interested in the
interaction in the long distance limit and hence the perturbative
expansion in terms of the long distance ($r \gg 1$) is enough for our
purpose.  Then, after some algebraic manipulation, it turns out that
the above formal result of the path integral is written as
\begin{align}
\prod^\infty_{n=0} \mathrm{Det}^{-1} M_n =
\prod^\infty_{n=0} {\det}^{-1} (\Delta_n P_n) \cdot
  {\det}^{-1} ( 1 - E_n ) \,,
\label{bdet2}
\end{align}
where the definition of $\Delta_n$ is given in Eq.~(\ref{deln}) and
the quantities inside the functional determinants are defined by
\begin{align}
P_n & \equiv 
  (\Delta_n-\frac{1}{3^2}) (\Delta_n-a_{n+})(\Delta_n-a_{n-})
  (\Delta_n-b_{n+})(\Delta_n-b_{n-})(\Delta_n-c_{n+})(\Delta_n-c_{n-}) \,,
\nonumber \\
E_n & \equiv
   \frac{r^2}{3^2} \frac{1}{P_n} (\Delta_n-b_{n+})(\Delta_n-b_{n-})
    \left[ (\Delta_n-a_{n+})(\Delta_n-a_{n-}) 
         + \frac{4}{3^2} \sigma (2n+1) (\Delta_n-\frac{1}{3^2})
    \right]
\nonumber \\
 & + \frac{2\sigma^2}{P_n} \Delta_n
    \left[ (2 \Delta_n-\frac{1}{3^2})(\Delta_n-\frac{1}{3^2})
          (\Delta_n-c_{n+})(\Delta_n-c_{n-})
          - \frac{r^2}{3^4} (14 \Delta_n - \frac{5}{3^2} )
    \right] \,,
\label{PE}
\end{align}
with
\begin{align}
a_{n\pm} & \equiv \frac{1}{18} \pm \frac{1}{3} 
  \sqrt{ r^2+\frac{1}{6^2} } \,,
\nonumber \\
b_{n\pm} & \equiv \frac{1}{18} \pm \frac{1}{3} 
         \sqrt{ r^2+ \sigma (2n+1)+\frac{1}{6^2} } \,,
\nonumber \\
c_{n\pm} & \equiv \frac{2}{3^2} \pm \frac{1}{3} 
\sqrt{ 3^2 r^2+ 4 \sigma (2n+1) + \frac{4}{3^2} } \,.
\label{abc}
\end{align}

We have considered the purely bosonic fluctuations.  The remaining
thing is the ghost part associated with the gauge fixing.  The
Lagrangian for the ghost part is
\begin{align}
L_G =  
\mathrm{Tr} \left[ \, 
 \dot{\bar{C}}^\dagger \dot{C} 
- \bar{C}^\dagger (r^2+ Q^2+P^2)  C
 +  \dot{\bar{C}} \dot{C}^{\dagger} 
- (r^2 +Q^2 +P^2) \bar{C} C^{\dagger} \,
\right] \, . 
\end{align}
The path integration is carried out by taking the same procedure used
for the bosonic part.  If we denote the modes of the ghost variables
$C$ and $\bar{C}$ as $c_n$ and $\bar{c}_n$ respectively, the
Lagrangian in terms of modes is obtained as
\begin{align}
L_G = 
&  \sum^{\infty}_{n=0}
\left[ \,
  \dot{\bar{c}}^*_n \dot{c}_n
+ \dot{\bar{c}}_n \dot{c}^*_n 
 - \left( r^2 + \sigma (2n+1) \right)
  ( \bar{c}^*_n c_n 
  + \bar{c}_n c^*_n ) \,
\right]\,,
\end{align}
whose path integration is straightforward and results in
\begin{align}
\prod^\infty_{n=0} \det{}^2 \Delta_n \,.
\label{gdet}
\end{align}

Let us now summarize the results and give the full expression obtained
in the bosonic and ghost part.  Eq.~(\ref{bdet1}) is the result of the
path integral for the matrix fields, $\Phi^8$ and $\Phi^9$.  For other
purely bosonic matrix fields, the path integral leads to
Eq.~(\ref{bdet2}).  As for the ghost part, we have obtained
Eq.~(\ref{gdet}).  The multiplication of these results gives
\begin{align}
\prod^\infty_{n=0}
\det{}^{-2}
( \Delta_n - \frac{1}{6^2} ) \cdot
\det{}^{-1} (\Delta_n^{-1} P_n) \cdot
\det{}^{-1} ( 1 - E_n ) \,,
\label{bdet}
\end{align}
which is the contribution to the one-loop effective action from the
bosonic and ghost part.  Here, we note that one half of the ghost
contribution remains in the final result.  This means that one half of
the unphysical gauge degrees of freedom is not canceled explicitly and
hidden in the bosonic contribution.  Of course, this will not give any
trouble. The effect of the gauge degrees of freedom is canceled
somehow by that of ghost in the actual evaluation of the functional
determinant.

\subsection{Fermionic contribution}

Having considered the bosonic and ghost fluctuations, let us turn to
the fermionic sector of the quadratic action.  The Lagrangian is
\begin{align}
L_F = 
&  \mathrm{Tr} \bigg[ i \chi^\dagger \dot{\chi} 
    - \frac{i}{4} \chi^\dagger \gamma^{123} \chi
    - r \chi^\dagger (\gamma^1 \cos(t/3) + \gamma^2 \sin(t/3) ) \chi
\notag \\
& + \chi^\dagger (\gamma^4 \cos(t/6) + \gamma^6 \sin(t/6) ) Q \chi
+ \chi^\dagger (\gamma^5 \cos(t/6) + \gamma^7 \sin(t/6) ) P \chi
\bigg] \, ,
\end{align}
where the matrix variable $\chi$ has been rescaled by a factor
$1/\sqrt{2}$.  As we have done in the previous subsection, we first go
to the frame, where background configuration becomes static one, by
taking the rotation
\begin{align}
\chi ~ \longrightarrow ~ \Lambda \chi \,,
\end{align}
where
$\Lambda = e^{-\frac{1}{6} t \gamma^{12}}e^{-\frac{1}{12} t \gamma^{46}} 
e^{-\frac{1}{12} t \gamma^{57}}$.
In this frame, the Lagrangian becomes
\begin{equation}
L_F = \mathrm{Tr} \, \chi^\dagger
\left[ \, 
   i \partial_t
     - \frac{i}{4}  \gamma^{123} 
     - r  \gamma^1  \,
     +  \gamma^4 Q + \gamma^5 P 
 -\frac{i}{12}  
  (2 \gamma^{12} + \gamma^{46} + \gamma^{57} ) \,
\right] \chi \,.
\label{Lf0}
\end{equation}

The above Lagrangian contains various products of gamma matrices,
which may lead to some complexity in practical calculation.  In order
to reduce the possible complexity, we first note the fact that $\chi$
is in $\mathbf{16}$ of $SO(9)$, that is, $\gamma_{(9)} \chi = \chi$
where $\gamma_{(9)} = \gamma^1 \gamma^2 \cdots \gamma^9$.  If we
consider the operator measuring the chirality in the $SO(6)$ symmetric
space as
$\gamma_{(6)}=\gamma^4\gamma^5\gamma^6\gamma^7\gamma^8\gamma^9$, we
see that $\gamma_{(9)} = \gamma^{123} \gamma_{(6)}$.  This shows that,
for a given eigenvalue of $\gamma_{(9)}$, the eigenvalue of
$\gamma^{123}$ is automatically determined by that of $\gamma_{(6)}$,
or vice versa.  In succession, because $\gamma_{(6)} = - \gamma^{46}
\gamma^{57} \gamma^{89}$, the chiralities in 4-6, 5-7, and 8-9 planes
determine the eigenvalue of $\gamma^{123}$.  At this point, we observe
that all the gamma matrices and their products in (\ref{Lf0}) commutes
with $\gamma^{89}$.  This means that the spinor components of $\chi$
with different eigenvalues of $\gamma^{89}$ do not couple in the
Lagrangian.  That is, if we split $\chi$ in terms of the chirality in
8-9 plane as
\begin{align}
\chi = \chi^{(+)}+\chi^{(-)} \,,
\end{align} 
where $\chi^{(s)}$ ($s=\pm$) has eight independent components and 
satisfies
\begin{align}
\gamma^{89} \chi^{(s)} = i s \chi^{(s)} \,,
\end{align}
then the Lagrangian is decomposed into two independent systems as 
follows:
\begin{align}
L_F = L_F^{(+)} + L_F^{(-)} \,.
\end{align}
Here, $L_F^{(s)}$ contains only $\chi^{(s)}$ and is given by
\begin{equation}
L_F^{(s)} = \mathrm{Tr} \,
\chi^{(s)\dagger} \left[ \, 
   i \partial_t 
     + \frac{s}{4} \gamma^{46} \gamma^{57}
     - r \gamma^1 
     +   \gamma^4 Q + \gamma^5 P
 -\frac{i}{12} 
  (2 \gamma^{12} + \gamma^{46} + \gamma^{57} ) \,
\right] \chi^{(s)}\,,
\label{Lfs}
\end{equation}
where we have used $\gamma^{123}=\gamma^{46} \gamma^{57}
\gamma^{89}$ as explained above.

As one may notice, splitting into two independent systems described by
$L_F^{(+)}$ and $L_F^{(-)}$ simplifies the problem, because each of
them is a system of spinor with eight independent components unlike
the original Lagrangian $L_F$ containing sixteen component spinor.  To
make more tractable form of each system, let us now split $\chi^{(s)}$
in terms of the chiralities in 1-2, 4-6, and 5-7 planes as
\begin{align}
\chi^{(s)} = \sum_{s_1, s_2, s_3 = \pm} \chi^{(s)}_{s_1s_2s_3} \,,
\end{align}
where $s_1$, $s_2$, and $s_3$ represent the eigenvalues of
$\gamma^{12}$, $\gamma^{46}$, and $\gamma^{57}$, respectively.  Then,
the action of $\gamma^{12}$ on $\chi^{(s)}_{s_1 s_2 s_3}$ is given by
$\gamma^{12} \chi^{(s)}_{s_1 s_2 s_3} = i s_1 \chi^{(s)}_{s_1 s_2
  s_3}$ and similarly for $\gamma^{46}$ and $\gamma^{57}$.

Besides the products of gamma matrices, the presence of $Q$ and $P$ in
the Lagrangian of Eq.~(\ref{Lfs}) leads to the mixing of modes with
different oscillator number $n$ when the spinor matrix
$\chi^{(s)}_{s_1s_2s_3}$ is expanded according to Eq.~(\ref{decom}).
As we have done in the bosonic case, such mixing problem is cured by
taking an appropriate unitary transformation and then newly defined
mode expansions for some variables.  We first consider the following
unitary transformation.
\begin{align}
\zeta_1^\pm &\equiv \frac{1}{\sqrt{2}} 
             (\gamma^4 \chi_{+++} \mp i \gamma^5 \chi_{+--}) \,,
\notag \\
\zeta_2^\pm &\equiv \frac{1}{\sqrt{2}} 
             (\gamma^4 \chi_{-+-} \mp i \gamma^5 \chi_{--+}) \,,
\notag \\
\zeta_3^\pm &\equiv \frac{1}{\sqrt{2}} 
             (\gamma^4 \chi_{-++} \mp i \gamma^5 \chi_{---}) \,,
\notag \\
\zeta_4^\pm &\equiv \frac{1}{\sqrt{2}} 
             (\gamma^4 \chi_{++-} \mp i \gamma^5 \chi_{+-+}) \,,
\end{align}
where the superscript $(s)$ in the spinor variables has been omitted
and its presence is implicit from now on.  This unitary transformation
is taken in such a way that the creation and annihilation operators
$a^\dagger$ and $a$ defined in Eq.~(\ref{osc-ca}) appear independently
in different terms.  After the transformation, we find that
$\zeta^\pm_1$ and $\zeta^\pm_4$ couple to each other as $-i
\sqrt{2\sigma} \zeta^{+\dagger}_1 a^\dagger \gamma^5 \zeta^-_4 +i
\sqrt{2\sigma} \zeta^{-\dagger}_1 a \gamma^5 \zeta^+_4$ and its
conjugation.  $\zeta^\pm_2$ and $\zeta^\pm_3$ have the similar
coupling.  Like the case of Eq.~(\ref{mode}), the structure of
couplings leads us to take the mode expansions for $\zeta_{2m}^\pm$
and $\zeta_{3m}^\pm$ as
\begin{align}
\zeta_2^\pm 
    = \sum^\infty_{n=\mp 1} \zeta_{2n}^\pm |n \pm 1 \rangle  \,, \quad
\zeta_4^\pm 
    = \sum^\infty_{n=\mp 1} \zeta_{4n}^\pm |n \pm 1 \rangle \,,
\end{align}
while $\zeta_1^\pm$ and $\zeta_3^\pm$ are taken to have the standard
mode expansion as in Eq.~(\ref{decom}).  Now, based on these mode
expansions, we see that the Lagrangian of Eq.~(\ref{Lfs}) does not
have any coupling between modes with different oscillator number, and
is written as
\begin{align}
L^{(s)}_F = \sum^\infty_{n=0} 
     Z^{(s)\dagger}_n F^{(s)}_n Z^{(s)}_n \,,
\label{slag}
\end{align}
where $Z^{(s)}_n=(\zeta^+_{1n},\zeta^-_{1n},\zeta^+_{2n},
\zeta^-_{2n},\zeta^+_{3n},\zeta^-_{3n},\zeta^+_{4n},
\zeta^-_{4n})^T$ and
\begin{align}
F^{(s)}_n = 
\begin{pmatrix}
K^{(s)}_1 & 0 & D & \Gamma_n  \\
0   & K^{(s)}_2 & \Gamma_n^\dagger & D \\
D & \Gamma_n & K^{(s)}_3 & 0 \\
\Gamma_n^\dagger & D& 0 & K^{(s)}_4
\end{pmatrix} \, .
\end{align}
The various quantities inside the matrix $F^{(s)}_n$ are $2\times 2$
matrices and defined by
\begin{align}
& K^{(s)}_1 =
\begin{pmatrix}
i \partial_t - \frac{1}{4}s +\frac{1}{6} & \frac{1}{6} \\
\frac{1}{6} & i \partial_t - \frac{1}{4} s +\frac{1}{6} 
\end{pmatrix} \, , \quad
K^{(s)}_2 =
\begin{pmatrix}
i \partial_t + \frac{1}{4}s - \frac{1}{6} & 0 \\
0 & i \partial_t + \frac{1}{4} s  - \frac{1}{6} 
\end{pmatrix} \, ,
\notag \\
& K^{(s)}_3 =
\begin{pmatrix}
i \partial_t - \frac{1}{4}s - \frac{1}{6} & \frac{1}{6} \\
\frac{1}{6} & i \partial_t - \frac{1}{4} s - \frac{1}{6} 
\end{pmatrix} \, , \quad
K^{(s)}_4 =
\begin{pmatrix}
i \partial_t + \frac{1}{4}s + \frac{1}{6} & 0 \\
0 & i \partial_t + \frac{1}{4} s  + \frac{1}{6} 
\end{pmatrix} \, ,
\end{align}
and
\begin{align}
\Gamma_n =
\begin{pmatrix}
0 & -i \sqrt{2 \sigma n} \gamma^5\\
i \sqrt{2 \sigma (n+1)} \gamma^5 & 0
\end{pmatrix} \, , \quad
D = r
\begin{pmatrix}
\gamma^1 & 0\\
0 & \gamma^1
\end{pmatrix} \, .
\end{align}  
We would like to note that, in writing the Lagrangian $L^{(s)}_F$ of
Eq.~(\ref{slag}), we have used the reasoning similar to that leading
to $L_B$ of Eq.~(\ref{Blag}), and symbolically identified
$\zeta^+_{2,-1}$ and $\zeta^+_{4,-1}$ with $\zeta^-_{2,0}$ and
$\zeta^-_{4,0}$ respectively.  Thus, the summation for $n$ starts from
$0$ for all modes.

The path integration for the Lagrangian $L^{(s)}_F$ is now evaluated
as
\begin{align}
\prod^\infty_{n=0} \mathrm{Det} F^{(s)}_n \,.
\end{align}
This is of course not the final form, and we should compute the matrix
determinant by exploiting the following matrix identity repeatedly.
\begin{align}
\begin{pmatrix} A & B\\ C & D \end{pmatrix} =
\begin{pmatrix} A & 0\\ C & 1 \end{pmatrix}
\begin{pmatrix} 1 & A^{-1} B\\ 0 & D-C A^{-1} B \end{pmatrix} \,.
\end{align}
After a little bit long manipulation, we see that $\mathrm{Det} F^{(s)}_n$ 
turns out to be written as
\begin{align}
\mathrm{Det} F^{(s)}_n = \det Q_n \cdot \det ( 1 - R^{(s)}_n ) \,,
\end{align}
where we have defined the following quantities
\begin{align}
Q_n = (\Delta_n - \hat{a}_{n+})(\Delta_n - \hat{a}_{n-})
      (\Delta_n - \hat{b}_{n+})(\Delta_n - \hat{b}_{n-}) \,,
\label{Qn}
\end{align}
\begin{gather}
\hat{a}_{n\pm} \equiv 
   \frac{5}{48} - \frac{1}{6}\sqrt{r^2 + \frac{1}{6^2}}
   \pm \alpha_{n-} \,, \quad
\hat{b}_{n\pm} \equiv 
\frac{5}{48} + \frac{1}{6}\sqrt{r^2 + \frac{1}{6^2}}
   \pm \alpha_{n+} \,,
\notag \\
\alpha_{n\pm} \equiv  \frac{1}{3}
 \sqrt{
   \frac{5}{2} r^2 + \frac{5}{4} \sigma (2n+1) 
   + \frac{59}{576} + 9 \sigma^2
   \pm \frac{5}{12} \sqrt{r^2 + \frac{1}{6^2}}
 } \,, 
\label{abhat}
\end{gather}
\begin{align}
R^{(s)}_n \equiv \frac{1}{Q_n}
&\bigg[
 -\frac{1}{6^2} \left( r^2 + \frac{1}{3 \cdot 4^2} \right) \partial_t^2
 +\frac{1}{9^2} \sigma (2n+1) 
     \left( -\partial_t^2 - r^2 + \frac{5}{12^2} \right)
\notag \\
& +\frac{4}{9} \sigma^2 \left(-\partial_t^2-r^2-\frac{1}{12^2} \right)
  + \frac{s}{2 \cdot 9^2} \sigma (2n+1) i \partial_t
\notag \\
& + \frac{s}{6}
   \left(\frac{1}{12} + \sqrt{r^2 + \frac{1}{6^2}} \right)
        (\Delta_n - \hat{a}_{n+})(\Delta_n - \hat{a}_{n-}) i \partial_t
\notag \\
& + \frac{s}{6}
   \left(\frac{1}{12} - \sqrt{r^2 + \frac{1}{6^2}} \right)
        (\Delta_n - \hat{b}_{n+})(\Delta_n - \hat{b}_{n-}) i \partial_t
 \bigg] \,.
\label{Rns}
\end{align}

By using the above result for the functional determinant, we can give
finally the full result of path integration for the fermionic
fluctuations as
\begin{align}
&\prod_{n=0}^\infty \mathrm{Det} F^{(+)}_n \cdot \mathrm{Det} F^{(-)}_n
\nonumber \\
=& \prod_{n=0}^\infty {\det}^2 Q_n \cdot \det ( 1 - R^{(+)}_n ) 
   \cdot \det ( 1 - R^{(-)}_n ) \,.
\label{fdet}
\end{align}

\section{Effective potential}
\label{epotential}

We are now ready to compute the effective potential by using the
formal path integral results obtained in the last section,
Eqs.~(\ref{bdet}) and (\ref{fdet}).\footnote{Almost all the
  computation in this section has been preformed by using {\it
    Mathematica}.}  The multiplication of these results gives $\exp (i
\Gamma_{\mathrm{eff}})$, where $\Gamma_{\mathrm{eff}}$ is the one-loop
effective action describing the interaction between the flat membrane
and the graviton.  The one-loop effective potential $V_{\mathrm{eff}}$
itself is related to the effective action via the relation
$\Gamma_{\mathrm{eff}} = - \int dt V_{\mathrm{eff}}$.

Each of the functional determinants in Eqs.~(\ref{bdet}) and
(\ref{fdet}) is composed of two parts.  One is the determinant whose
argument has the factorized form and another with the argument of
fractional form.  We first consider the former one, and compute its
contribution to the effective potential in the limit of large
distance.  Then the relevant functional determinant is given by the
multiplication of $\prod^\infty_{n=0} \det{}^{-2} ( \Delta_n -
\frac{1}{6^2} ) \cdot \det{}^{-1} (\Delta_n^{-1} P_n)$ and
$\prod_{n=0}^\infty {\det}^2 Q_n$.  The calculation of this can be
done without much difficulty, and the potential is read off as
\begin{align}
&\sum^\infty_{n=0}
\bigg[ 
2 \sqrt{m^2_n + \frac{1}{6^2}} + \sqrt{m^2_n + \frac{1}{3^2}}
+\sqrt{m^2_n + a_{n+}}+\sqrt{m^2_n + a_{n-}}
\notag \\
& +\sqrt{m^2_n + b_{n+}}+\sqrt{m^2_n + b_{n-}}
+\sqrt{m^2_n + c_{n+}}+\sqrt{m^2_n + c_{n-}}
-\sqrt{m^2_n}
\notag \\
& - 2 \sqrt{m^2_n + \hat{a}_{n+}} - 2 \sqrt{m^2_n + \hat{a}_{n-}}
 - 2 \sqrt{m^2_n + \hat{b}_{n+}} - 2 \sqrt{m^2_n + \hat{b}_{n-}} \,
\bigg] \,,
\end{align}
where $m^2_n \equiv r^2 + \sigma (2n+1)$ and other quantities have
been already defined in Eqs.~(\ref{abc}) and (\ref{abhat}).  This
expression contains the infinite sum over $n$, which may lead to the
issue of convergence.  However, the large $n$ behavior of the summand
can be shown to be of the order of $n^{-3/2}$, and thus the summation
is well-defined and convergent.  The sum over $n$ itself can be
performed by adopting the Euler-Maclaurin formula
\begin{align}
\sum^\infty_{n=0} f(n) = \int^\infty_0 dx f(x) + \frac{1}{2} f(0)
-\frac{1}{12} f'(0) + \frac{1}{720}f'''(0) + \dots
\label{emf}
\end{align}
which is valid when $f$ and its derivatives vanish at infinity.  After
the summation, if we expand the resulting potential in terms of large
$r$, we obtain
\begin{align}
\frac{\sigma}{r} - \frac{67}{15552} \frac{1}{\sigma r} 
+ \frac{1}{216} \frac{\sigma}{r^3} 
- \frac{215}{373248} \frac{1}{\sigma r^3}
+ \mathcal{O} \left( \frac{1}{r^5}\right) \,.
\label{pot0}
\end{align}
We see that this potential is quite ridiculous in physical sense
because of the terms which are inversely proportional to $\sigma$.
Vanishing $\sigma$ means the absence of the flat membrane in the
background configuration, and hence it is expected that there is no
potential.  But, the above potential diverges when $\sigma$ is set to
zero.  Therefore, for getting sensible result, the terms which are
inversely proportional to $\sigma$ must be canceled by contributions
from other functional determinants.  As it should be, we will see that
such terms are absent at the final result.

We now turn to the effective potential which comes from the functional
determinants with fractional argument. Let us first consider the
contribution from the bosonic part, that is, $\prod^\infty_{n=0}
{\det}^{-1} ( 1 - E_n )$ in Eq.~(\ref{bdet2}).  From the identity
\begin{align}
\det{}^a (1+ A) = \exp [ \, a \mathrm{tr} \ln (1+A) \, ]
= \exp[ \, a \mathrm{tr} A - a \mathrm{tr} A^2/2 + \dots ]
\end{align}
where $\mathrm{tr}$ is the functional trace, we see that, through
the simple power counting, the leading
contribution to the effective potential at large distance is 
$i \sum^\infty_{n=0} \mathrm{tr} E_n$.  The trace calculation of this
is transformed to an integration in momentum space.  After
evaluating the integration, the Euler-Maclaurin formula (\ref{emf})
and the expansion in terms of large $r$ then lead us to have the
following contribution to the effective potential.
\begin{align}
i \sum^\infty_{n=0} \int^\infty_{-\infty} 
\frac{d \omega}{2\pi} E_n =
-\frac{\sigma}{r} + \frac{23}{3888} \frac{1}{\sigma r}
+ \frac{1}{216} \frac{\sigma}{r^3} 
+ \frac{521}{933120} \frac{1}{\sigma r^3}
+ \mathcal{O} \left( \frac{1}{r^5}\right) \,,
\label{pot-bos}
\end{align}
where $\omega$ is the conjugate variable of time $t$, and
$\partial_t^2$ inside $E_n$ is understood to be replaced by
$-\omega^2$.  This is the leading order result in the large distance
limit.  As a next step, one may consider the next to leading order
contribution given by $(i/2) \sum^\infty_{n=0} \mathrm{tr} (E_n)^2$.
However, this and higher order contributions are turned out to give at
most $\mathcal{O} ( 1/r^5 )$, and hence they are not our concerns.

As for the fermionic part, the functional determinant with fraction
argument is given by $\prod_{n=0}^\infty \det ( 1 - R^{(+)}_n ) \cdot
\det ( 1 - R^{(-)}_n )$ of Eq.~(\ref{fdet}).  The procedure of
evaluating this is the same with that taken in the bosonic case.  The
leading order contribution to the effective potential in the large
distance limit is obtained as
\begin{align}
- i \sum^\infty_{n=0} \int^\infty_{-\infty} 
\frac{d\omega}{2\pi} \left[ R^{(+)}_n + R^{(-)}_n \right] =
- \frac{25}{15552} \frac{1}{\sigma r}
- \frac{1}{27} \frac{\sigma}{r^3} 
+ \frac{1}{38880} \frac{1}{\sigma r^3}
+ \mathcal{O} \left( \frac{1}{r^5}\right) \,.
\label{pot-fer1}
\end{align}
We note that, as can be noticed from the definition of $R^{(s)}_n$ in
Eq.~(\ref{Rns}), the terms depending on the sign $s$ do not enter at
the leading order because of the cancellation between terms with
opposite signs.  Those terms contribute at the next to leading order.
If we calculate the contribution from the next to leading order, then
we get
\begin{align}
- \frac{i}{2} \sum^\infty_{n=0} \int^\infty_{-\infty} 
\frac{d\omega}{2\pi} \left[ (R^{(+)}_n)^2 + (R^{(-)}_n)^2 \right] =
- \frac{1}{124416} \frac{1}{\sigma r^3}
+ \mathcal{O} \left( \frac{1}{r^5}\right) \,.
\label{pot-fer2}
\end{align}
For higher order contributions, the power counting tells us that the
potential is of the order of $\mathcal{O} ( 1/r^5 )$.  Thus, there is
no need to consider higher orders in the large distance limit.

Up to now, we have obtained all the necessary results for giving the
one-loop effective potential between graviton and flat membrane in the
large distance limit.  If we sum up the results obtained in
Eqs.~(\ref{pot0}), (\ref{pot-bos}), (\ref{pot-fer1}) and
(\ref{pot-fer2}), then the final expression of the one-loop effective
potential in the large distance limit becomes
\begin{align}
V_{\mathrm{eff}} = - \frac{1}{36} \frac{\sigma}{r^3} + 
\mathcal{O} \left( \frac{1}{r^5}\right) \,.
\label{effpot}
\end{align}
We see that the leading order interaction is of $r^{-3}$ type.
Because there is no other contribution giving this kind of interaction
term, it can be concluded that the leading interaction of the
effective potential is one-loop exact.  As an important remark, we
also see that the potential is completely sensible in a way that it
does not contain terms which are inversely proportional to $\sigma$.
All such terms have been canceled exactly with each other.

It should be noted that the effective potential (\ref{effpot}) has the
same type with that of the interaction potential \cite{Shin:2008ep}
between fuzzy sphere (giant graviton) and flat membrane in plane-wave
matrix model.  In the context of the DLCQ M-theory in the flat
spacetime, the interaction between graviton and flat membrane is
usually expected to be of $r^{-5}$ type, as explicitly illustrated in
\cite{Aharony:1996bh}.  In \cite{Shin:2008ep}, the result that the
leading order interaction at large distance is of $r^{-3}$ type rather
than $r^{-5}$ type has been interpreted as the one due to the smearing
of flat membrane in two extra spatial dimensions.  The potential
(\ref{effpot}) obtained in this paper supports this interpretation and
provides one more evidence about it.

\section{Conclusion}
\label{finalsec}

Motivated by the previous observation and interpretation in the study
of the interaction between fuzzy sphere membrane and flat membrane, we
have considered the configuration composed of one point-like graviton
and one flat membrane, and investigated the interaction between them
in the context of plane-wave matrix model.

At the one-loop level, the effective potential between graviton and
flat membrane has been obtained, and its leading order interaction in
the large distance limit has been shown to be of $r^{-3}$ type.  In
\cite{Shin:2008ep}, this type of interaction rather than $r^{-5}$ type
has been interpreted as the delocalization or smearing effect due to
the configuration of the flat membrane which spans and spins in four
dimensional space.  Our final result (\ref{effpot}) agrees well with
this interpretation.  Furthermore, it provides one more evidence for
the smearing effect due to the configuration of flat membrane in
plane-wave background.

\section*{Acknowledgments}

This work was supported by the National Research Foundation of Korea
(NRF) grant funded by the Korea government (MEST) through the Center
for Quantum Spacetime (CQUeST) of Sogang University with grant number
2005-0049409.  The work of H.S. was also supported by the National
Research Foundation of Korea (NRF) grant funded by the Korean
government with the grant No.~KRF-2008-331-C00071 and
R01-2008-000-21026-0.

\end{document}